\documentstyle[12pt,twoside,fleqn,espcrc1,epsfig]{article}
 
\newcommand{\AmS}{{\protect\the\textfont2
  A\kern-.1667em\lower.5ex\hbox{M}\kern-.125emS}}

\title{Strangelet Searches in High Energy Heavy Ion Collisions}

\author{K.N. Barish\address{ 
                            Department of Physics and Astronomy, University of
                            California at Los Angeles, 405 Hilgard Ave, Los Angeles, CA 90095}
        for the E864 Collaboration\address{
                            Ames Lab -- Univ Bari -- BNL -- UCLA -- Iowa State -- Univ Mass --
                            MIT -- Penn State -- Purdue -- USMA -- Vanderbilt -- Wayne State -- Yale}
}

\begin{document}
% typeset front matter
\maketitle

\begin{abstract}
In this contribution I concentrate on the recent results from experiment E864 at the BNL-AGS.
E864's recent analysis have achieved sensitivities of approximately $3\times 10^{-8}$ 
per 10\% central interaction for the production of
charged strangelets and the first analysis for neutral strangelets is near completion.
I put the results in the context of coalescence and quark gluon plasma strangelet production models.
\end{abstract}

%
%  Uncomment out if preprint format required
%
%\pacs{00.00, 20.00, 42.10}
%\maketitle

\section{Introduction}

Relativistic heavy ion collisions provide an environment that can produce systems 
with a large baryon number and strangeness content in a small volume. About 400 baryons interact and more than
20 $s\bar{s}$ pairs are produced in a central Au+Pb event at the AGS.

Strange quark matter (SQM) states are $A>1$ hadrons which contain
strange quarks and within which the constituent quarks are allowed to
move freely. The presence of strange quarks
lowers the Fermi energy levels compared to that of ordinary quark
matter (QM -- which contains only up and down quarks). This may mean 
that SQM is more stable than QM, despite the addition of the more
massive strange quark. 

Calculations~\cite{chin_kerman,farhi_jaffe,gilson_jaffe} based on QCD and
the phenomological bag model suggest that SQM may be meta-stable or
even completely stable for a wide range of bag model parameter values,
which are otherwise unconstrained by past experiments on nuclear
matter. Excluding shell effects, all of the calculations contain the feature
that SQM systems become more stable as the baryon number $A$ increases.

In this contribution, I concentrate on the search for SQM with
$A<100$ (often referred to as {\it strangelets}) conducted by experiment
E864 at the BNL-AGS. 

\section{Experimental strangelet searches}

The production probability for strangelet formation is likely different for different colliding systems 
and energies. Collisions with heavier systems likely produce a more probable environment for strangelet
production. Collisions at AGS energies ($\approx$$11.6~A~GeV/c$ for Au beams)
are likely more favorable for coalescence production due to 
more limited expansion of the system. In contrast, QGP production may be more prominent at 
CERN energies ($\approx$$158~A~GeV/c$ for Pb beams).
However, the strangeness distillation mechanism (the process by which a QGP may evolve into a strangeness
rich system) may be weaker at CERN energies due to the lower baryon density.

The experimental search for strangelets in heavy ion collisions began with 
experiment E814 at BNL-AGS with Si beams~\cite{E814} (Si+Au). Subsequently, experiments
E858 (Si+Au)~\cite{E858}, E878 (Au+Au)~\cite{E878}, and E886 (Au+Pt)~\cite{E886} conducted 
strangelet searches. At the CERN-SPS strangelet searches at higher energies have been conducted by
experiment NA52 (using S and Pb beams). 
None of these experiments have seen evidence for strangelet production.

Because the experiments measure only a limited region of phase space 
a production model is need to translate null
measurements into production limits. All of the experiments use a similar model in which
the rapidity ($y$) and transverse momentum ($p_{_t}$) are taken to be uncorrelated: 
\begin{equation}
  {d^2N\over{dydp_{_t}}} \propto p_{_t} e^{-{2 p_{_t} \over
  <p_{_t}>}}e^{-{(y-y_{_{cm}})^2 \over 2\sigma_{_y}^2}} $$
 \label{eq:diffxs}
\end{equation}
where $<p_{_t}>$ is the mean transverse momentum, $y_{_{cm}}=1.6$ is the
center-of-mass rapidity for $11.6~GeV/c$ Au+Pb collisions, and $\sigma_{_y}$ is the standard deviation
of the rapidity distribution of the strangelet.  

The experimental searches, with the exception of E814 and E864, employ focusing spectrometers. 
These spectrometers, which consist of quadrupole as well as dipole magnets, have acceptance
for a fixed rigidity ($R=p/Z$, where $p$ is momentum and $Z$ charge). They have
very small acceptances but have little background and can sample much beam to achieve high sensitivities.
However, the disadvantage is that they are strongly dependent upon the production model assumed. 
E886 and E878 have maximum rigidity settings of $2~GeV/c$ and $20~GeV/c$ respectively. These rigidity 
limitations mean the experiments do not measure near mid-rapidity for higher mass strangelets 
(i.e., $m>10~Gev/c^2$).

By contrast, experiment E864 uses an open geometry spectrometer. E864 has now completed its second round 
of strangelet searches, and is in the process of analyzing further data. The recent results from E864 
now represent the most sensitive search to date for most masses and production models. E864 is also the
first experiment with the capability to search for neutral strangelets.

\section{The E864 Spectrometer}

\begin{figure}[t]
   \centering
   \epsfxsize=5.0in
   \leavevmode
   \epsffile[101 354 538 585]{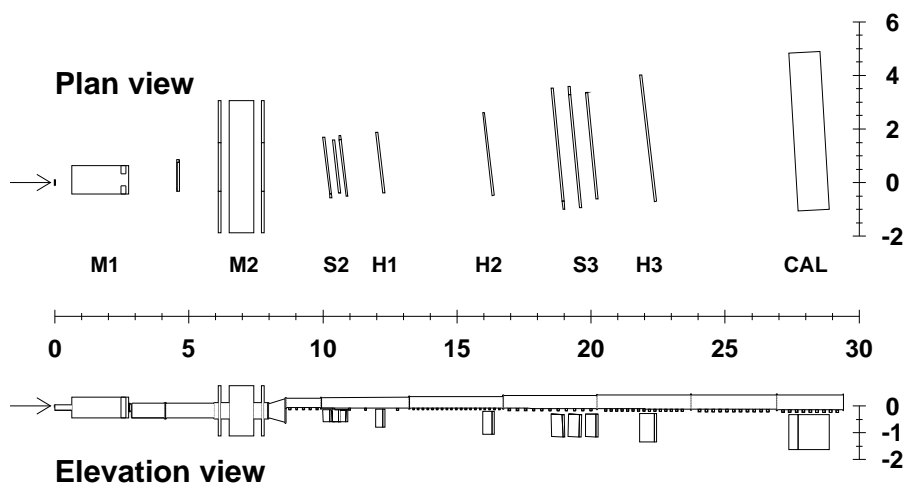}
   \caption{The 1995 configuration of the E864 spectrometer. In the
      plan view, the downstream vacuum chamber is not shown. M1 and M2 are
      dipole analyzing magnets, S2, and S3 are straw tube arrays, H1-H3 are
      scintillator hodoscopes, and CAL is a hadronic calorimeter.  The
      horizontal and vertical scales are in meters.}
   \label{fig:spec}
\end{figure}

A schematic diagram of the E864 spectrometer is shown in Fig.~\ref{fig:spec}. 
The open geometry design of the E864 spectrometer translates into a
large geometric acceptance and leads to an efficient search. 
The spectrometer consists of two dipole magnets,
three segmented hodoscope planes of scintillation counters (206 vertical slats), 
two stations of straw tube arrays (4 mm diameter tubes), and a
lead/scintillating fiber hadronic calorimeter (754 towers)~\cite{e864_cal}.
The spectrometer is designed to measure particles around center-of-mass rapidity ($y_{cm}=1.6$) 
because this is where we expect strangelet production to be peaked.
For the strangelet analysis, E864 searches over an approximate rapidity
range from $y=1.1$ to  $y=2.1$.

The charge ($Z$) of a
particle that traverses the spectrometer is measured using pulse height
information from the 3 hodoscope walls.  Its rigidity ($R$) is derived from the
target position and downstream slope and position
of the particle's track as
measured by the straw tube and hodoscope detectors.  The particle's velocity
is measured using timing information from the hodoscopes, giving the
relativistic quantities $\beta$ and $\gamma$.  The particle's
mass ($m$) is then reconstructed as $m = {R \over \gamma \beta}Z$.
The calorimeter's time and energy information is used to confirm
above measurements, reject potential backgrounds, and make measurements for neutral
particles.

In order to achieve high sensitivities, the detector is able to 
record 1000-2000 events every 4 seconds (AGS spill). E864 is able to 
sample several million Au+Pb interactions per second 
with the implementation of two hardware triggers: a centrality trigger
and a high-mass trigger (or Late-Energy Trigger -- LET).

The centrality trigger uses a segmented scintillation
multiplicity counter which is located $13$~cm downstream the target.
The multiplicity counter system, which measures the interaction's products
within an angular range of 16.6$^\circ$ and 45$^\circ$ with respect to the
incident beam direction, provides a rough measure of the impact parameter, or
centrality, of the reaction~\cite{beam}.  For this analysis, 
events with the 10\% highest pulse heights, or roughly the 10\% most
central events are selected.

The calorimeter is used to provide a high-level hardware trigger 
which selects high-mass objects. The time and energy from each tower are used to
make a rough mass measurement ($m=E/(\gamma-1)$). A programmable
lookup table to select high-mass objects is loaded for each tower.
With this trigger, an enhancement of $>50$ for high-mass
objects in the recorded data sample was achieved. 

\section{Recorded Data}

The E864 spectrometer was commissioned during the Fall 1994 run of the
experiment. Only 1/4 of the
calorimeter was completed and there was no high-mass trigger. 
For the 1995 experimental run, the calorimeter and high-mass trigger
were fully implemented.  

A summary of the data recorded by experiment E864 is shown in Table~\ref{tab:e864data}. The ``+1.5T'' data
is optimized for the positive strangelet search, while the ``$-$0.75T'' data is optimized for the negative
strangelet search. All data was taken with a 11.6~A~GeV/c Au beam.
\begin{table}
\caption{Summary of E864 Recorded Data}
 \label{tab:e864data}
 \centering
\begin{tabular}{||c|c|c|c|c||} \hline
Data             & Target    & Central            & LET            & Sampled                      \\
                 &           & Events Recorded   & Enhancement     & Central Collisions           \\ \hline
                 &           &                   &                 &                              \\
1994\ \  ``+1.5T''   & 10\% Pb & $\ 26.5\times 10^6$ & N/A             &  $\ 2.7\times 10^7$          \\
1995\ \  ``+1.5T''   & 30\% Pb & $\ 27.0\times 10^6$ & 55              &  $\ 1.5\times 10^9$          \\
1995 ``--0.75T''& 30\% Pb & $\ 85.5\times 10^6$ & 55              &  $\ 4.7\times 10^9$          \\
1996\ \  ``+1.5T''   & 60\% Pt & $185.0\times 10^6$  & 73              &  $13.5\times 10^9$           \\
1996 ``--0.75T''& 60\% Pt & $217.0\times 10^6$  & 77              &  $16.7\times 10^9$           \\ \hline
\end{tabular}
\end{table}

\section{Results from E864 Strangelet Searches}

\subsection{Charged Strangelet Analysis}

From the 1994 analysis~\cite{barish,E864},
E864 observed 25 $Z=+1$ strangelet candidates with masses from 
$10-200$~GeV/c$^2$.
However, from GEANT simulations carried out during the design phase of the
experiment, a class of $Z=+1$
backgrounds was expected for the 1994 configuration of the detector from
neutrons which undergo inclusive charge exchange reactions $(n+A\to p+X)$ 
after the second magnet but before the first detector.
The protons can travel in a similar direction to the original
neutron, which does not bend in the magnetic field. Therefore, the
reconstructed tracks resulting from this
process appear to have a high rigidity and thus a high mass.
E864 performed studies to compare their observations to their expectations that
the backgrounds were due to charge exchange and
concluded that the high mass candidates are consistent with the expected
background.  
Thus, E864 placed upper limits on strangelet production.

\begin{figure}[t]
   \centering
   \epsfysize=3.5in
   \leavevmode
   \epsffile{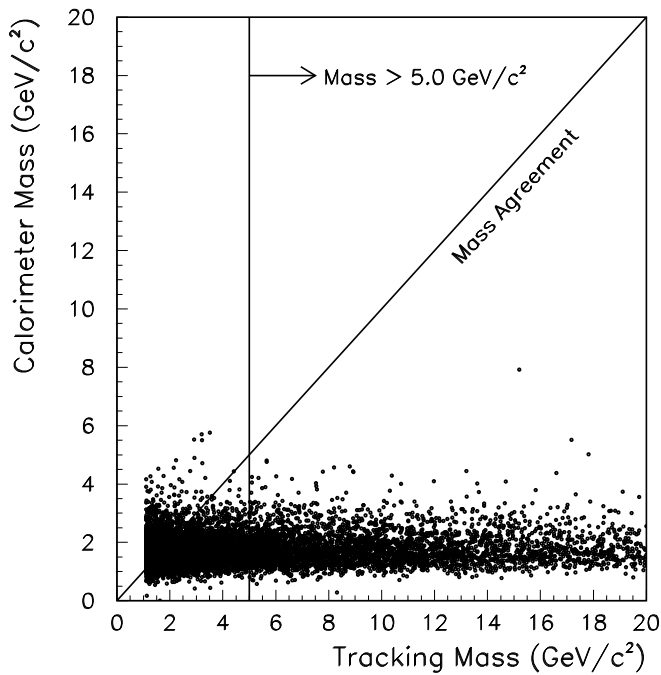}
   \caption{A scatter plot of $Z=-1$ high-mass candidates' tracking mass
       versus calorimeter reconstructed mass. The data represents the
       complete ``$-$0.75T'' 1995 data sample after calorimeter contamination
       cuts and track cuts have been applied. }
   \label{fig:neg_mass_mass}
\end{figure}

In addition to the tracking system, the full calorimeter
is used for the 1995 analysis. Fig.~\ref{fig:neg_mass_mass}
illustrates the power of the calorimeter. 
Plotted is the mass
calculated from the energy deposited in the calorimeter and time at
the calorimeter $(m=E_{dep}/(\gamma-1))$ versus the associated mass 
reconstructed by the  tracking system for the ``$-$0.75T'' 1995 data. 
Cuts have been applied to ensure that the calorimeter energy is not 
contaminated from other showers, and tracking $\chi^2$ cuts to fits in
the bend plane, vertical plane, and time have been 
applied to ensure good tracks. The time at the calorimeter  
is derived from a fit to the hodoscope times because this provides a
more accurate time measurement than using the time derived from the
calorimeter. Note that the target point is not
included in this fit to allow for tracks which do
not originate from the target. From this figure, it can be seen that
the tracks which reconstruct with a high mass from the tracking
detectors do not have a correspondingly high calorimeter mass. Indeed,
when agreement between the calorimeter and tracking
chambers is imposed, all of the high-mass candidates are eliminated.

For the ``+1.5T'' positive strangelet search~\cite{E864,coe},
no candidates were observed for $Z=+1$ with $m>5\ {\rm GeV/c}^2$, or for $Z=+2$  
with $m>6\ {\rm GeV/c}^2$. 

For the ``$-$0.75T'' strangelet search~\cite{E864,nagle}, 
no candidates were observed for $Z=-1,-2$ with $m>5\ {\rm GeV/c}^2$.
Because of our large geometric acceptance, E864 was also able to search
for positive strangelets with this spectrometer setting. These limits
are combined with the ``+1.5T'' limits to produce our final sensitivities.

E864 does observe clear peaks in their mass distributions for
$Z=1$ objects such as $p$, $d$, $t$, $K^{-}$, and $\bar{p}$, and
$Z=2$ objects such as $^3$He, $^4$He, and $^6$He. For the 1995 data sample
E864 sees over 50 $^6$He nuclei measured within $\pm 0.6$ units of mid-rapidity.

\subsection{Neutral Strangelet Analysis}

The production of neutral strange quark matter in heavy ion collisions is 
previously unconstrained by experimental measurements.
E864 is able to be sensitive to neutral strangelets with use of the hadronic 
calorimeter. 

The hodoscope and straw tube tracking chambers are used to identify charged particles
entering the calorimeter. The calorimeter showers which these correspond to are rejected. 
The energy and timing information from the calorimeter is then used to
reconstruct the mass of neutral particles. 

\begin{figure}[t]
   \centering
   \epsfysize=3.5in
   \leavevmode
   \epsffile{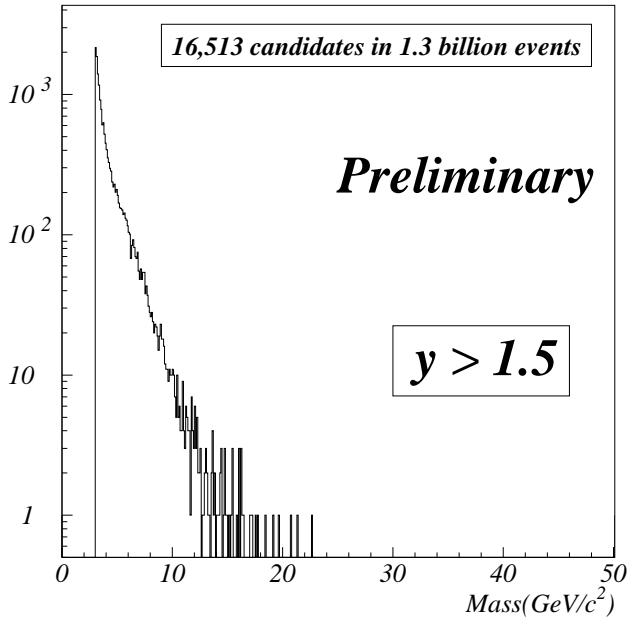}
   \caption{Neutral candidates after charged particle subtraction.}
   \label{fig:neut_mass}
\end{figure}

Unlike the charged strangelet analysis, in which the rejection power of the tracking system is utilized, 
the neutral analysis is expected to be limited by backgrounds. Three source of expected background are
overlapping neutron showers, charged particles which have not been identified
by the tracking system, and ``double interaction'' events in which a second interaction occurs
after the original interaction and deposits energy in the calorimeter. 
An example mass distribution from the 1995 ``+1.5T'' data sample is shown in 
Fig.~\ref{fig:neut_mass}.~\cite{muhnoz}.
A detailed analysis to optimize the charged particle subtraction and 
to understand the neutral mass distribution is underway.

E864 will have sensitivity for low mass neutral states, such as the quark
alpha state (a single quark bag consisting of 6 up, 6 down, and 
6 strange quarks), as well as higher mass states.

\subsection{Limits on Strangelet Production}

\begin{figure}[t]
   \centering
   \epsfysize=3.5in
   \leavevmode
   \epsffile{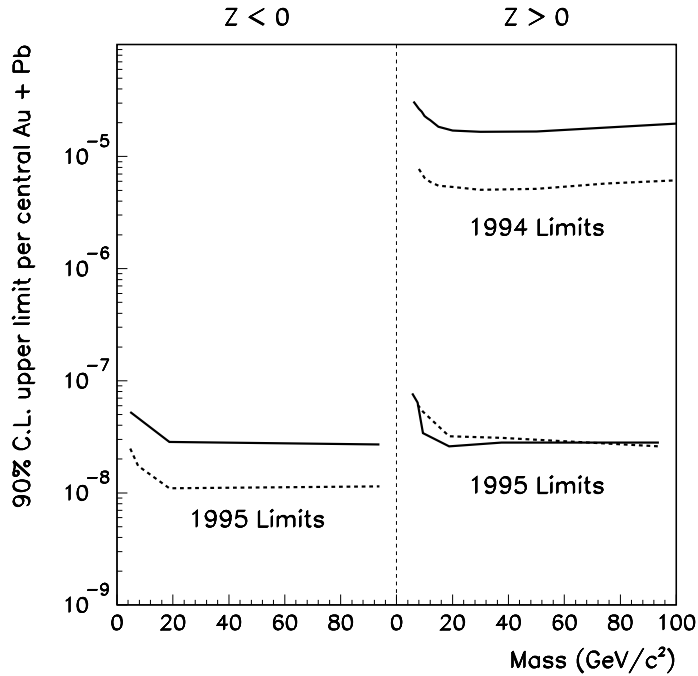}
   \caption{90\% confidence level limits in 10\% central collisions  
       for $Z=\pm 1$ and $Z= \pm 2$ strangelet production with
       lifetimes in excess of 40~ns. The solid lines correspond to $Z=\pm 1$, while the
       dashed lines correspond to $Z=\pm 2$.}
   \label{fig:limits_hyp97}
\end{figure}

The results from the 1994 and 1995 charged strangelet searches are shown in 
Fig.~\ref{fig:limits_hyp97}.
The solid lines correspond to $Z=\pm 1$, while the dashed lines correspond to $Z=\pm 2$.

E864's upper limits are nearly flat as a function of mass, owing to the
large acceptance of the spectrometer.  These limits are only mildly sensitive
to changes in
Eq.~\ref{eq:diffxs} for the same reason.  For example, if the rapidity width of
strangelet production were taken to be $y = 0.5/\sqrt{A}$, the E864
curves in Fig.~\ref{fig:limits_hyp97} would be lower (give better limits) by less
than a factor of 2.

A comparison between the 1994 and 1995 limits show an improvement 
of well over two orders of magnitude. This is possible because the
calorimeter provides additional background rejection, and the high
mass trigger allows us to sample more events.

E864 has estimated a preliminary sensitivity for neutral strangelets in the absence of background 
(e.g. $m\;_{\sim}^{>}\;30$) from the 1995 ``+1.5T'' data.
The preliminary sensitivity is approximately $1\times 10^{-7}$ per 10\% central Au+Pb interaction.

\section{Interpretation of Strangelet Limits}

Ideally, one would like to be able to translate sensitivity limits
into constraints on the bag model parameters. However, with the
current state of the theories regarding strangelet production and
stability this is not possible.
But the limits can be compared to the available predictions
from coalescence and quark-gluon plasma models. 

In order to compare E864's limits with the predictions of some models,  
the difference between the measurements in minimum bias and 10\%
central events must be accounted for.  Since strangelet production requires the ingredients of many
baryons and many units of strangeness, it is expected that the
probability to produce strangelets is greater in central
interactions than in peripheral interactions.
Ref.~\cite{baltz} predicts that roughly 50\% of the production
probability for an $A=6$ hypernucleus with 2 units of strangeness 
resides in the 10\% most central events. Thus, for comparison we assume that 50\% of 
the strangelet production resides in the top 10\% most central
collisions by dividing the central limits by a factor of 5.

\subsection{Coalescence Models}

\begin{figure}[t]
   \centering
   \epsfysize=3.5in
   \leavevmode
   \epsffile{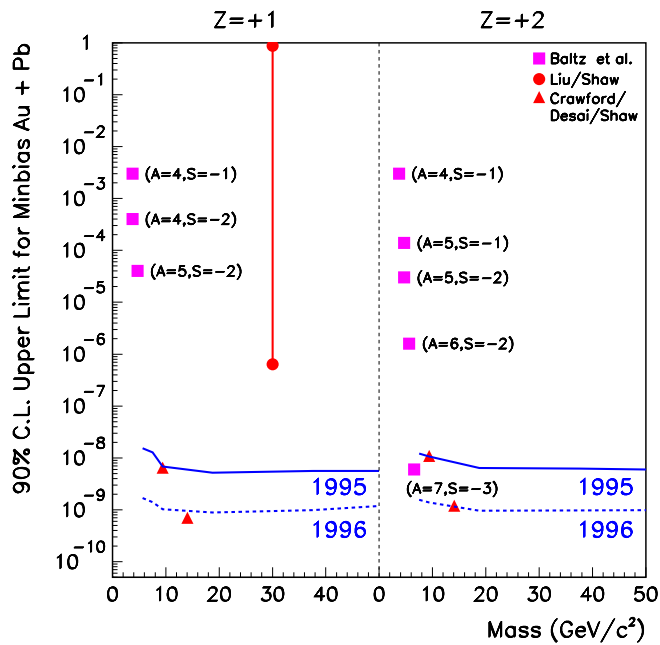}
   \caption{Comparison between E864's limits and predictions from some
      models. For the comparison, the limits are divided by a factor
      of 5 to convert the central limits to minbias (see text). The line
      labeled 1995 are the limits from the ``+1.5T'' 1995 run. The dotted
      lines labeled 1996 are the projected sensitivities from the 1996 run
      given the amount of data recorded and known LET enhancement. The
      solid squares are predictions based on the coalescence model from
      Ref.~\protect\cite{baltz}. The solid circles with the line are
      the range of predictions based on QGP formation from from 
      Ref.~\protect\cite{liu}. The solid triangles are predictions
      based on QGP formation from Ref.~\protect\cite{crawford}.}
   \label{fig:compare_hyp97}
\end{figure}

Heavy ion collisions, in which many hyperons are produced in a small region 
of phase space, is an environment in which strangelets may be produced by
coalescence. In this model, the ingredients of a
strangelet (baryon number, strangeness, and charge), in the form of
ordinary baryons, can fuse to form a strangelet. Predictions from
Ref.~\cite{baltz} are shown in Fig.~\ref{fig:compare_hyp97}. 
One might
conclude that E864 is beginning to address coalescence predictions for
low masses. However, the reader must be cautioned that when taking the
predictions from Ref.~\cite{baltz}, one can infer that this model
over predicts the experimentally observed yields of $^3$He and 
$^4$He presented in Ref.~\cite{pope}. Further work is required to interpret the
sensitivity in the context of coalescence models.

\subsection{Quark-Gluon Plasma Models}

In another class of models, a strangelet could result from the
formation of a quark-gluon plasma (QGP). These production models require
(1) that a QGP is produced, and (2) that the QGP charges up with strange
quarks compared to non-strange quarks and cools down into a strangelet.

For the first step, Kapusta {\em et al.}~\cite{kapusta} propose that most
collisions at AGS energies produce superheated hadronic matter.
However, in rare events, a droplet of QGP is nucleated, converting most
of the matter to plasma.  They calculate the probability that thermal
fluctuations in a superheated hadronic gas will produce a plasma
droplet, and that this droplet will be large enough to overcome its
surface free energy to grow. They estimate the probability for this 
to occur to be on the order of once every 10 to once every 1000 central
collisions. A high mass strangelet would be an experimental 
signature from which we could infer the formation of a QGP even if it
is was a rare occurrence. For the second step, Greiner {\em et
al.}~\cite{greiner} suggest that a QGP would likely evolve into
a strangelet via the process of strangeness distillation (assuming
that strangelets are meta-stable). In this process, once a QGP is
formed, it will cool by emitting mesons. ${\bar s}$ quarks will
preferentially find $u$ and $d$ quarks to form $K$ mesons in a baryon
rich plasma, such as would be formed at AGS energies. The QGP charges
up with strange quarks relative to anti-strange quarks when the mesons are
emitted by the plasma. After the system cools down, a strangelet may be
formed. These strangelets would likely be massive because the size of
the produced QGP could be large, and the large strangelets are
postulated to be the most stable.

The data restricts these processes at the 90\% confidence level
approximately as follows:
\begin{equation}
  {\rm BR}({\rm Au+Pb} \to {\rm QGP}) \times {\rm BR}({\rm QGP} \to 
  {\rm Strangelet}) < 2.5 \times 10^{-8}
  \label{eq:z1lim}
\end{equation}
where BR(Au+Pb$\to$QGP) is the probability for a central Au+Pb collision at 
$11.6~A~GeV/c$ to produce a QGP, and BR(QGP$\to$Strangelet) is the probability
of the QGP to cool into a strangelet.
Equation \ref{eq:z1lim} is approximately valid for strangelets with
masses between 10 and 100~GeV/c$^2$.  

Previously, there have been numerical predictions on the production
rate of strangelets assuming QGP formation. Predictions by Crawford
{\em et al.}~\cite{crawford} are the solid triangles in
Fig.~\ref{fig:compare_hyp97}. The original predictions were for Si+Au
collisions, but they provided a formula which was applied to derive
their Au+Au predictions. E864 is beginning to achieve sensitivities which
address their predictions.

Liu and Shaw~\cite{liu} made the first predictions for strangelet
formation. Their predictions are for Si beams, and are quite model
dependent. However, one might expect higher production rates with
heavier Au+Pb collisions. Their range of predictions, shown in 
Fig.~\ref{fig:compare_hyp97} by a solid line connecting two circles, 
are ruled out by E864's measurements.

\section{Prospects for the Future}

E864 will finish with the 1995 neutral analysis and continue with the the
1996 neutral and charged analysis. Given the amount of
data recorded during the 1996 run,  the positive strangelet sensitivity should improve by roughly 
a factor of 10 (see the dashed line in Fig.~\ref{fig:compare_hyp97}) and the
negative strangelet sensitivity should improve by by roughly a factor of 3.5.
The 1996 data is also being used to search for strangelets with 
$|Z|>2$. E864 has already seen $Z=3$ objects such as $^6$Li from small sample of
the 1996 data set. Schaffner {\em et al.}~\cite{jurgen} have recently made calculations indicating that
low mass strangelets with large negative charges are more likely to be seen.
The sensitivity to neutral strangelets will also improve significantly with the 
1996 data. In addition to the increase in data volume, improvements were made to
help suppress the ``double interaction'' background which was seen in the 1995 analysis.

The current round of strangelet experiments are sensitive only to strangelets with
lifetimes in excess of $\approx 40~ns$. Strangelets are
theorized to become more stable as they become more massive. 
Therefore, it is likely that even if
stable strangelets exist, they have masses larger than can be produced
via the coalescence mechanism. However, if the lighter strangelets are
meta-stable, it would be possible to study them if the experiments were
sensitive to particles with lifetimes similar to hyperons.
Hopefully experiments will be built which are sensitive to 
these shorter lived strangelets.

\section{Acknowledgments}

E864 would like to thank the BNL AGS staff for providing the Au beam for the experiment.
This work was supported by grants from the U.S. Department of Energy's
High Energy and Nuclear Physics Divisions, the U.S. National Science
Foundation, and the Istituto Nazionale di Fisica Nucleare of Italy.

\end{document}